# USE OF ROOT IN VEHICULAR ACCIDENT RECONSTRUCTION

Bob Scurlock, Ph.D.

*Department of Physics, University of Florida, Gainesville, Florida*

**Introduction**

The purpose of this article is to introduce the reader to the free data analysis software package ROOT. ROOT is superior to, and more flexible than, many other commonly used data analysis tools, including most spreadsheet applications. In this article, simple examples are given to demonstrate how ROOT may used to complement one's accident reconstruction analyses. To aid the new user, example ROOT application scripts, as well as tips and tricks, can be found at the free educational website: SoftwareForAccidentReconstruction.com.

**ROOT**

ROOT is the physics analysis tool currently favored by most particle physicists around the world [1]. ROOT was created by Rene Brun and Fons Rademakers at the European Organization for Nuclear Research (CERN). It was created in anticipation of the startup of the Large Hadron Collider (LHC) – a particle collider machine. The ROOT package originally dates back to 1994. At its essence, ROOT is a collection of C++ (object-oriented) libraries along with a C++ interpreter called CINT. The user takes advantage of this extensive library by creating various "objects" that contain all of the needed functionality to perform calculations. The user has the option to execute commands interactively with the ROOT command-line interface, through runtime-executed scripts, or by creating compiled code, which links against the ROOT library.

With ROOT, one can generate Monte Carlo data, create multi-dimensional histograms, create graphs, fit data, perform matrix and vector manipulations, generate visual aids, create platform independent GUIs and much more. ROOT is a constant work in progress, whereby feedback from its users is critical to its evolution. ROOT has undergone intense quality control and validation over the past decade, and has been instrumental in performing high precision measurements of phenomena at the smallest scale. It is used to store, retrieve, and analyze enormous amounts of data created by some of the world's most precise measurement devices.

While ROOT can seem overwhelming to new users because of its incredible breadth of functionality, there are many gentle introductions available on the web. In fact, it is quite common for ROOT to act as a novice's introduction to the C++ programming language. Once the user becomes comfortable with the basic ideas of object-oriented programming, ROOT becomes quite easy to use. As one becomes more proficient in the usage of ROOT, the analysis possibilities become endless, and the line between data analysis and data production becomes seamless. ROOT is currently used in fields of study including subatomic particle physics, astrophysics, cosmology, quantitative finance, data-mining, medical sciences, and of course, vehicular accident reconstruction, and is available for free.

ROOT is supported on many platforms including Windows, MacOS, and Unix. ROOT code is completely portable between different operating systems. Installation information and documentation can be found on the world wide web (which was also invented at CERN) at http://root.cern.ch.

**Scripting with ROOT**

ROOT enables its user to quickly prototype usable software by its ability to interpret code as a series of commands issued during runtime. This mode of work is often referred to as scripting, and utilizes ROOT's CINT runtime command interpreter. In addition to speedy development cycles, scripting is a very advantageous way of working as CINT is quite forgiving for poor programming etiquette. The penalty of using a runtime interpreter is of course, that more CPU cycles are required to finish execution; however, depending on the use case, this may not be of particular importance to the programmer.

**Simple Scripting Example**

The connection between accident reconstruction and ROOT becomes clear when one considers typical use cases of other computational software packages used in accident reconstruction, such as spreadsheet applications. While spreadsheet applications certainly have their advantages in that development time can be rapid, scalability becomes a challenge as one is often constrained by time-consuming mouse commands, as well as cumbersome data entry by hand. While it may take slightly more time to enter values into two data arrays in order to create a graph within ROOT, one may quickly generalize the task by scripting. This may then allow the user to easily understand how variations to input data influence the output of a given calculation. For example, suppose one wanted to understand the braking distance required to bring a vehicle to full stop as a function of pre-braking velocity, assuming a constant braking coefficient of friction. This is a common type of calculation reconstructionists often need to perform. Using energy conservation, one simply arrives at the usual second-order polynomial result as a first approximation: $d = v_i^2/(2\mu g)$. In a spreadsheet, one would need to create multiple rows and columns in order to see what incremental variations in $\mu$ bring to the estimate. In ROOT, this can simply be accomplished by utilizing a two-dimensional histogram and a standard Monte Carlo analysis approach. One simply issues the series of commands shown in Figure 1. For those knowledgeable with C++ programming, the syntax should look familiar. The example script "instantiates" particular objects from a few ROOT classes. For example, the "TH2F" class contains everything one would need to create a two-dimensional histogram, which one could think of as a two-dimensional array of numbers with floating point precision. A random number generator object "TRandom" is instantiated, and used to generate the random input data to the model, using *Gaussian* and *Uniform* probability density functions (pdf's). ROOT classes such as those used for creating histograms, contain (member) functions to draw, manipulate, and perform calculations on their own (member) data, which is held in memory. In this example, an instance of this TH2F histogram class is created to store the results of velocity versus braking distance calculations; that is, the histogram is used to contain the results from all Monte Carlo trials. Member functions of the very same histogram class are then used to create plots of its data. When plotted, the resulting TH2F may be drawn as a scatter plot, lego plot, contour plot, or surface plot, depending on the user input.

**Graphical Data Analysis**

Another great feature of ROOT is its ability to graphically represent data. Indeed there are many ways to parse, manipulate, project, and understand data within ROOT. For example, the graphical output of our simple braking distance script is shown in Figure 2. Here the data is represented as a temperature-style scatter plot, where the color indicates the number of Monte Carlo trials that resulted in an ordered pair ($v_i$, $d$). Because the data is binned in a two-dimensional grid, the color actually indicates the number trials resulting in data which fall within the range $v_i \pm \Delta v_i/2$ and $r \pm \Delta r/2$, where each bin corresponds to a cell $\Delta v_i \times \Delta r$ in size.

One of the advantages of a Monte Carlo style analysis lies in the ability to make selection-cuts to examine particular regions of input "parameter-space," typically with the ultimate goal assigning probability or likelihood for a set of input values to fall within a particular range. This allows the user to specify results within a particular confidence limit. In our example, if we suppose that the scene evidence or witness statements indicate that a vehicle required between 10 and 20 feet to come to a full stop, then we may wish to select only those trials that result in outcomes consistent with that evidence. The two-dimensional histogram class within ROOT allows one to project data onto the x-axis (y-axis), with the option of selecting only that data which lies in a band of bins along the y-axis (x-axis). In our case, we can choose to look at the resulting velocity distribution, given that we only what to see trials in which the braking distance was between 10 and 20 feet. In ROOT, this can be accomplished with a single command, which uses the projection member function of the TH2F class. Figure 3 shows the output of this command. The left side shows our original plot, where we zoomed into the region of interest by simply using ROOT's mouse zoom feature. On the right is the initial velocity distribution, given that we have selected only those trials that result in braking distances between 10 and 20 feet. The resulting distribution has a Gaussian shape. Using the simple fit command, we obtain a Gaussian curve centered at 25.8 fps with a 4.5 fps standard deviation. This of course, implies that our 95% C.L. initial velocity range is 16.8 – 34.8 fps given our constraints on braking distance between 10 – 20 feet. Rather than using the projection member function of the TH2F class, we could have explicitly placed our selection criteria within the for-loop of the script. This approach generally allows the user to place multiple constraints on the output data for a more sophisticated analysis. This represents a significant improvement over using spreadsheet applications as ROOT naturally lends itself to this style of analysis.

Although this trivial example represents a simple use of ROOT, it is meant to demonstrate how a straightforward, yet very informative, analysis could be accomplished in less than 20 lines of code. The number of tools available to the ROOT user is extensive. Indeed, we could have created a visualization of the vehicle braking, which could also run interactively within ROOT. This visualization could include representations of force vectors, running clocks, and other features. ROOT has incredible flexibility.

Figure 4 shows an example use of three-dimensional histogram used to visualize a pedestrian impact simulated with a simple projectile approximation. Figure 5 shows the relationship between impact-velocity and throw distance for such a pedestrian impact. The results for Monte Carlo trials are again stored in 2-dimensional histogram form, and drawn as a temperature plot. Various model curves are superimposed using the "TF1" 1-D function class [2,3,4,5]. Taking advantage of ROOT's "TGraph" class, Figure 6 shows the corresponding $1\sigma$ and $2\sigma$ bands for the $\chi^2$ distribution for all models. Selecting the region about throw distance ~ 35 m, we obtain a best impact velocity solution of approximately 63 kph, as shown in Figure 7.

Another example of ROOT's flexibility is shown in Figure 8 where a single frame from a primitive animation of Articulated Total Body (ATB) output data is shown [6]. In this visualization, the head, neck, upper-, mid-, and lower-torso segments are displayed using ROOT's "TEllipse" class. The forces imparted to those segments are represented using the "TArrow" class. In addition to the animated 2-D projected view of the body segments, the position, velocity, acceleration, and force versus time relations can simultaneously be displayed in order to help the user understand the underlying dynamics of the simulation, which determine the occupant motion. This type of application would be extremely difficult to achieve with a spreadsheet tool.

A final example of data visualization is shown in Figure 9. Here we see the output of a damage-based analysis script, where the CRASH3 algorithm is applied to obtain closing-velocity and delta-V values as a function of crush depth and crush width [7]. Another 2-D histogram is used in this case, with the "TH2F::SetBinContent()" method used to set the data values along the Z-axis rather than the "TH2F::Fill()" method. In this way, the content of each histogram bin is set to a single unique value. The Z-axis value (velocity) is the read by the color scale. Such a display allows the reconstructionist to quickly obtain a feel for the magnitude of a given collision event, even before having a precise set of vehicle crush measurements at hand. An alternative to this 2-D histogram representation could have been created by drawing a surface graph using the "TGraph2D" class (Figure 10).

Scripting with ROOT allows the user to combine subroutines developed during the course of his or her reconstruction career in a very natural way. For example, one could use the previously mentioned CRASH3 subroutine to obtain a likely range of pre-impact velocity values for a given collision. This range of values could be represented as an input pdf to be used in the script shown in Figure 1. This would then modify the final distribution of braking distance versus pre-impact velocity based on the constraints obtained by the damage-based analysis. Over time, one develops a library of such subroutines which can be reused in many combinations to suit the requirements of a given subject accident. This represents yet another advantage of ROOT over spreadsheet applications.

**Interactive Scripting Versus Compiling**

The example script presented in Figure 1 was not compiled, but rather interpreted by the CINT interpreter, which is built into ROOT. CINT interprets and executes user code at runtime. After execution, all objects declared within the script are kept in memory (unless explicitly deleted) for the user to continuing working with on the ROOT prompt. In this way, one can for example reformat the look and feel of a graph defined within a script, after runtime. This mode of programming allows for very rapid development and debugging. Once code is refined through multiple debug cycles, one typically adapts the code to compile within ROOT. When ROOT is allowed to compile user code, it creates permanent object files such that source code is recompiled only when changes are detected. Compiled code generally runs a factor of 10 times faster than interpreted code, and thus can become very important as the scale and complexity of one's software increases over time.

**Data Storage**

Another advantage to working with the ROOT software package is in the ability to blend the data production and analysis steps in a nearly seamless way. One can generate large amounts of data for a study, and then store the data in ROOT's own tree data structure format. This is a powerful feature that allows one to store variable length arrays of an arbitrary number of variables, all connected by a common index (this could be Monte Carlo trial number or a simulation time step for example). This allows the user to study correlations across many Monte Carlo trials, without having to regenerate all data each time the analysis is updated or revisited. A data file generated this way can then be reopened at a later time, and analyzed using custom data analysis classes, which ROOT automatically templates specifically to work with the user's own data tree file.

In addition to the ability to store correlated data sets, one can also store the graphical representations of the data. Histograms and graphs can be stored in these ROOT files. These files can be reopened, and the histogram and graph data contents can be reloaded and drawn interactively. Line colors can be changed, as well as every other visible feature, well after the original data was produced. One's data can be kept persistent indefinitely if needed.

ROOT is excellent for parsing data from flat form ASCII data files. This type of data can be converted into the ROOT tree format, or for simplicity, stored as a collection of one-dimensional histograms, where bin number corresponds to the ASCII file row number. Storing data in the ROOT tree format allows one to take advantage of ROOT's data browser (Figure 11), where one can point-and-click to immediately visualize data. This can be used for example to quickly visualize output data from vehicle dynamics simulation software packages.

**Conclusion**

In this article, simple examples illustrating a few features of the powerful data analysis tool ROOT are presented. The focus was on Monte Carlo style studies; however, uncertainties could have been estimated through finite differences, closed-form expressions, or by other methods, all using ROOT. ROOT provides means by which one can quickly program calculations and simulations within custom subroutines and graphically visualize their results. The subroutines can be reused and combined over time, as the users needs evolve. Beginners typically take one month to become comfortable with basic scripting and plotting techniques, but quickly make progress customizing their code to perform complex analyses. ROOT is proven, powerful, and free.

*Bob Scurlock is a Research Associate at the University of Florida, Department of Physics, a member of the Compact Muon Solenoid Experiment (CMS) at CERN, and owner of Scurlock Scientific Services, LLC. He can be reached at: BobScurlockPhD@gmail.com.*

```
{
  TRandom ranGen;
  TH2F myHisto("Histo_VivsR","Range Plot",100,0,100,10000,0,1000);
  int Ntrials = 1E6;
  for (int i=0; i<Ntrials; i++)
    {
      float g = 32.174;
      float vi = ranGen.Uniform(0,100);
      float mu = -1;
      while (mu<0)
         mu = ranGen.Gaus(0.7, 0.2);
      float d = vi*vi/(2*mu*g);
      myHisto.Fill(vi, d);
    }
  myHisto.GetXaxis()->SetTitle("Initial Velocity (fps)");
  myHisto.GetYaxis()->SetTitle("Braking Distance (ft)");
  myHisto.Draw("COLZ");
}
```

Figure 1: Example ROOT (interpreted) script. This simple code will generate a scatter plot showing pre-braking velocity versus braking distance to stop.

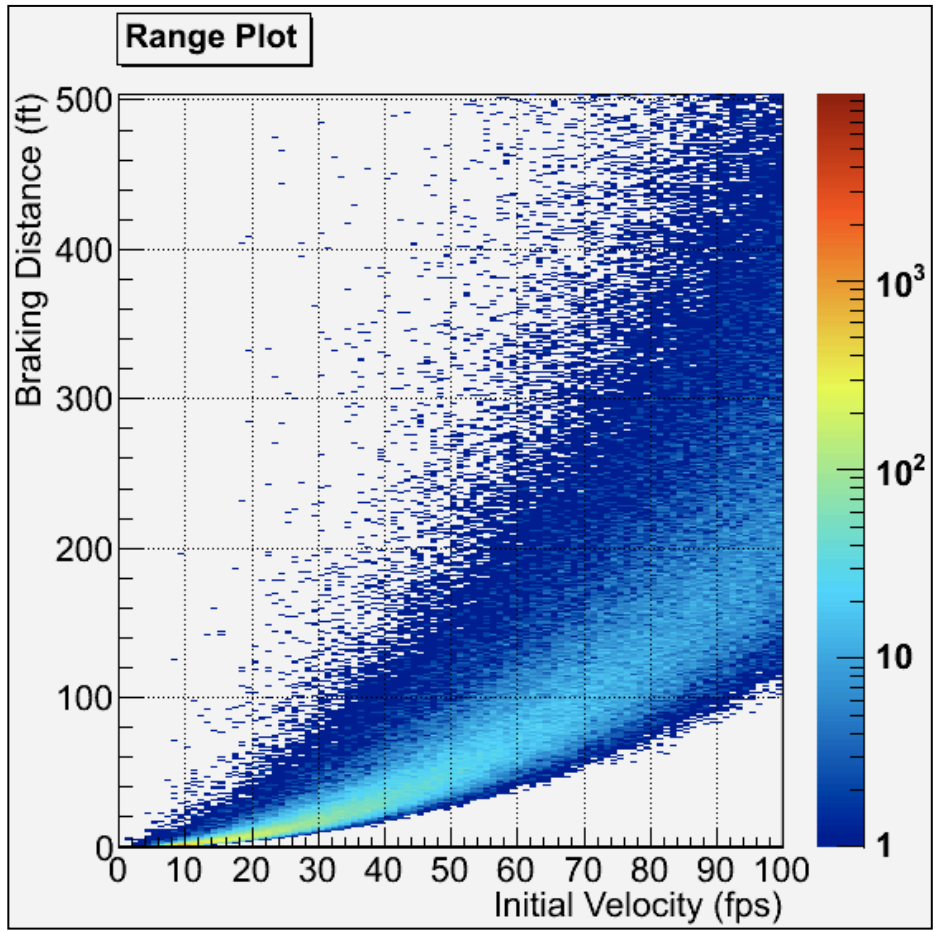

Figure 2: Two-dimensional temperature-style histogram generated by example script. The relative intensity for Monte Carlo trials to result in a particular outcome

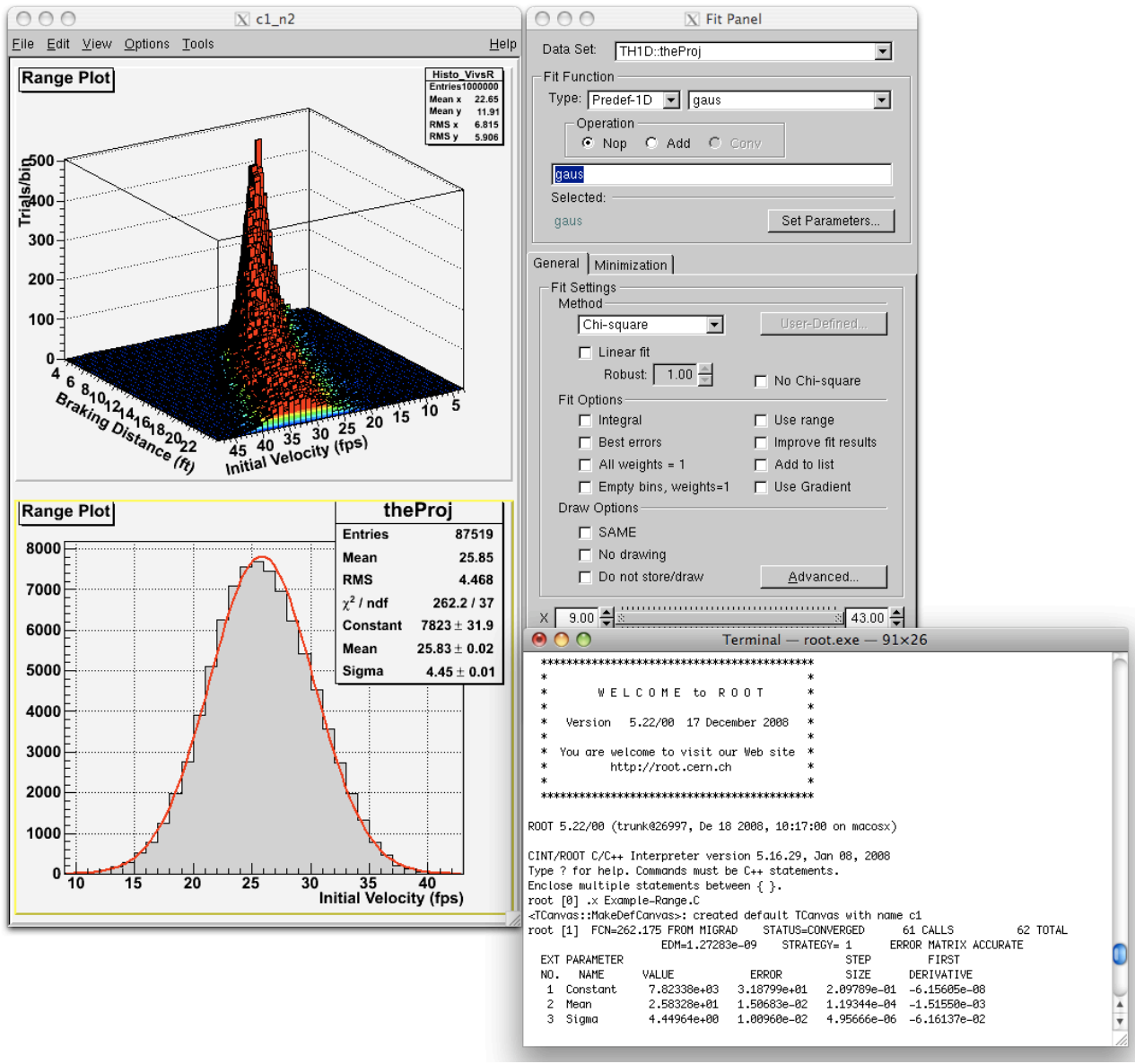

**Figure 3:** Upper left shows lego plot of braking distance versus initial velocity distribution. Bottom right left shows fit to data within 10 - 20 feet braking distance range. Right shows ROOT's fit panel menu3.

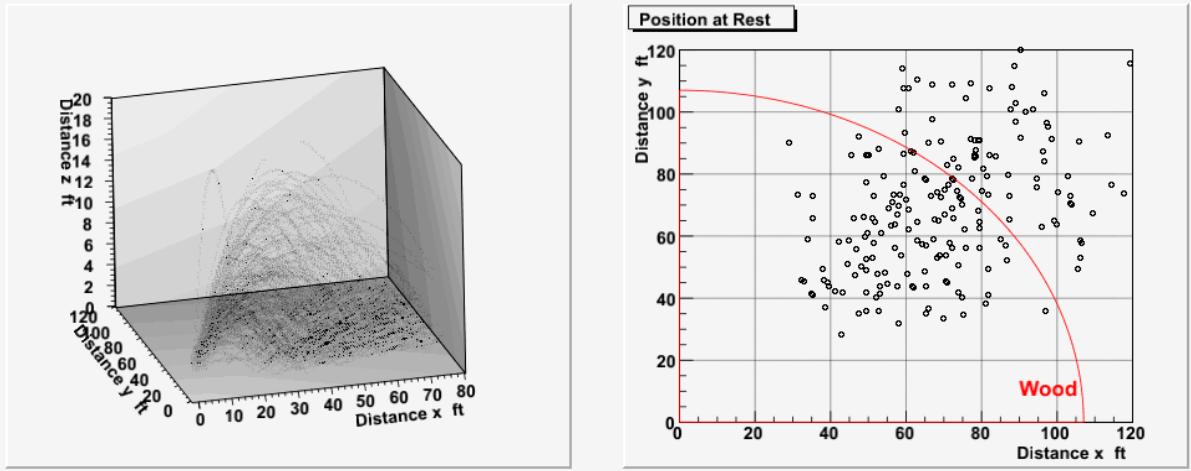

**Figure 4:** (Left) Visualization of Monte Carlo trials of a simple projectile simulation. (Right) Points-of-rest for Monte Carlo trials. Wood and Simms model result is also shown.

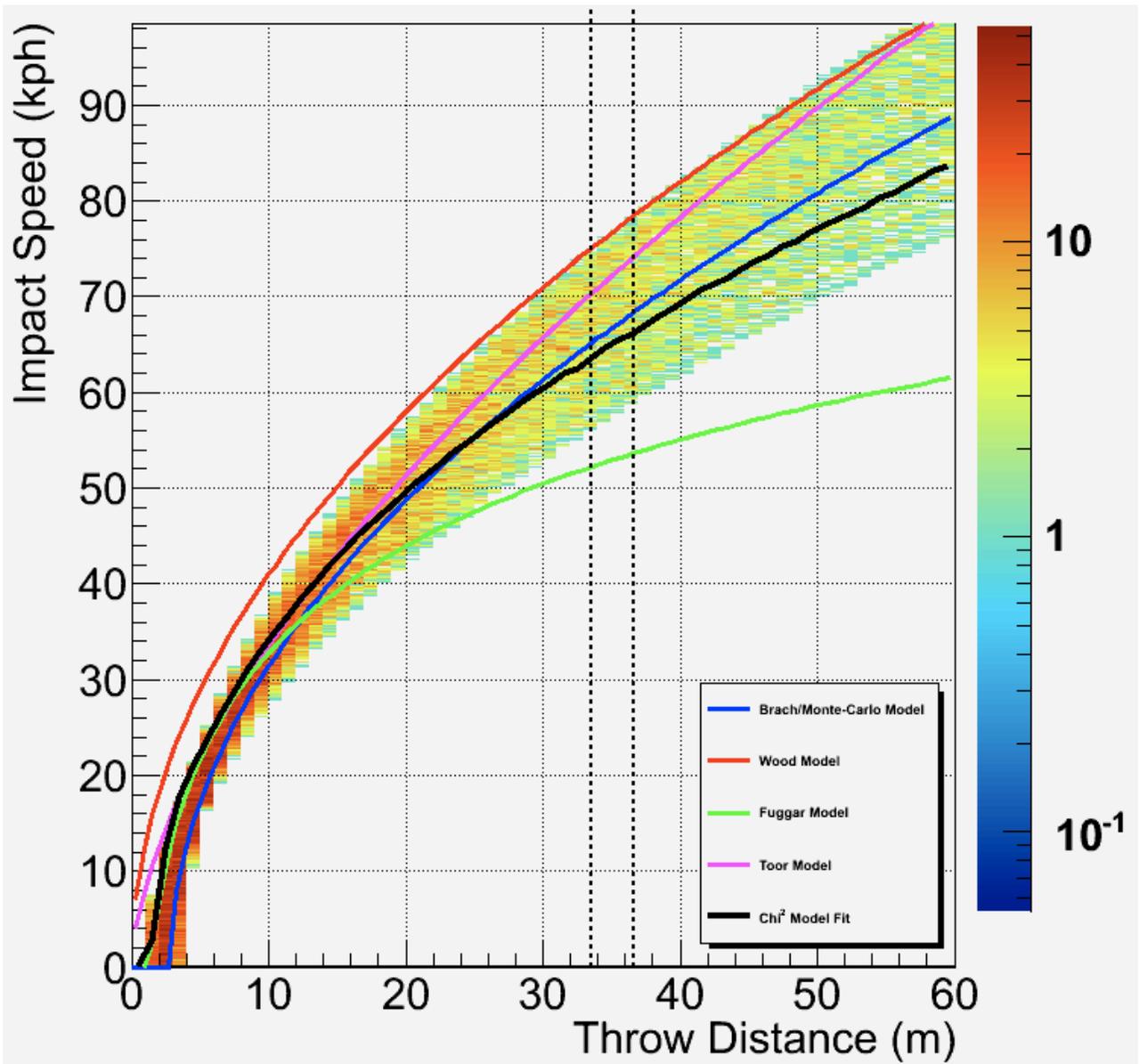

Figure 5: Pedestrian impact speed versus throw distances for Monte Carlo trials. Various throw-model curves are also shown.

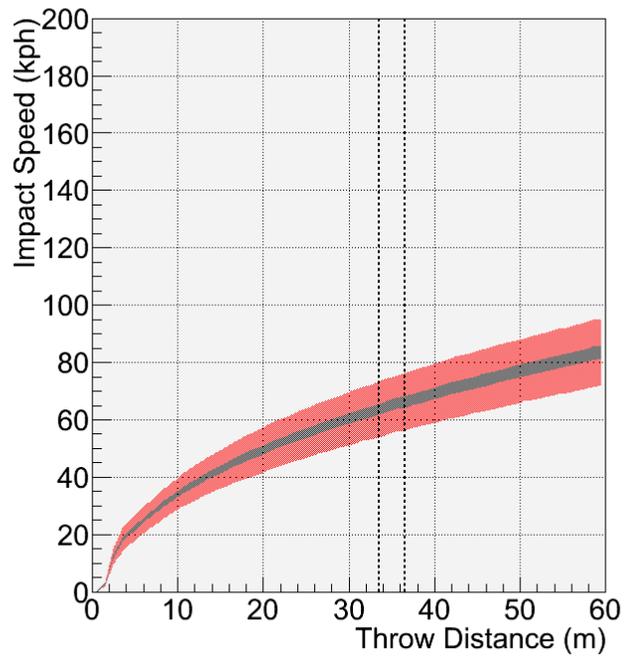

Figure 6: 1σ and 2σ bands for the $\chi^2$ distribution over all pedestrian impact models6.

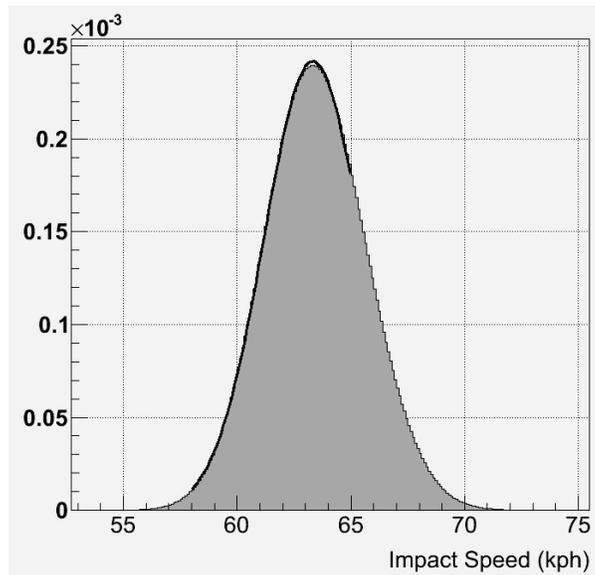

Figure 7: Selecting the region about 35 m from the $\chi^2$ distribution shown above, we see the most likely impact speed centered at 63 kph.

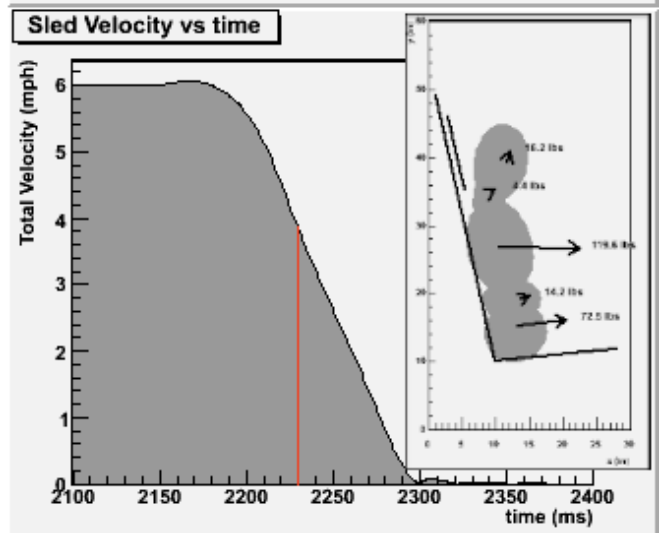

Figure 8: Single frame from a ROOT animated visualization of Articulated Total Body data for a rear impact collision.

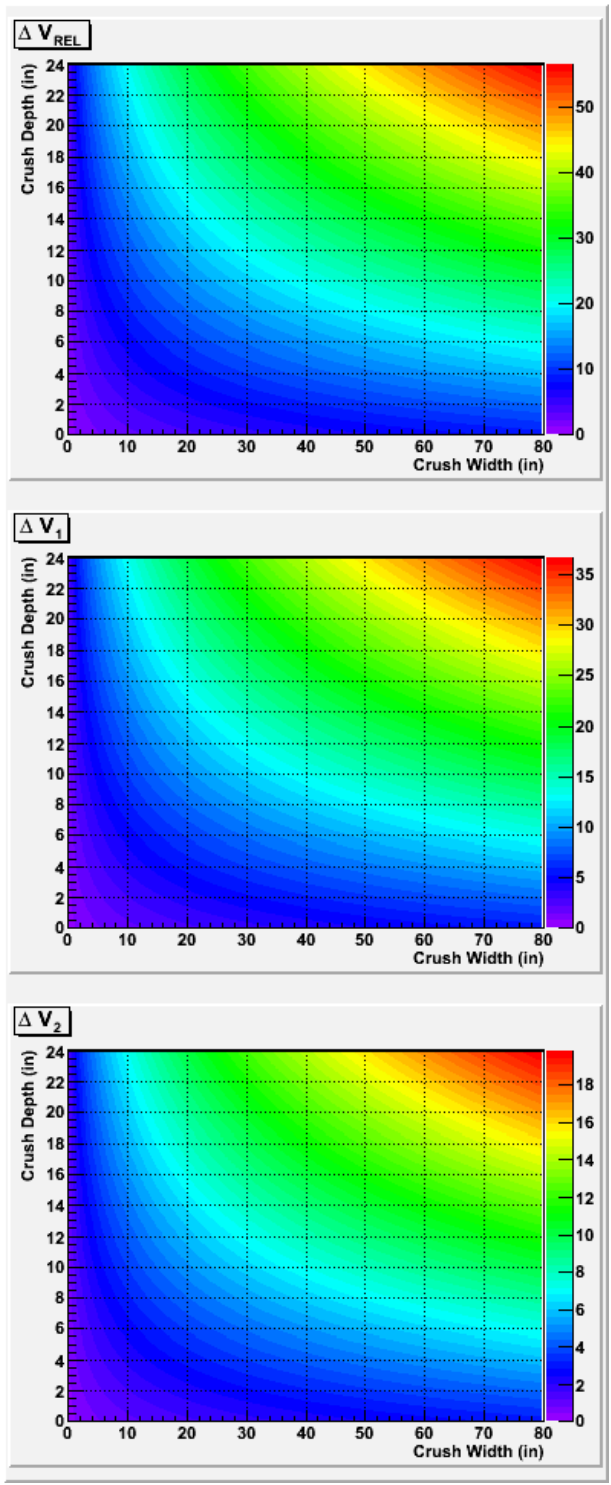

Figure 9: Output of crush damage based analysis script. Here the closing-velocity and delta-V values are shown versus crush depth and crush width.

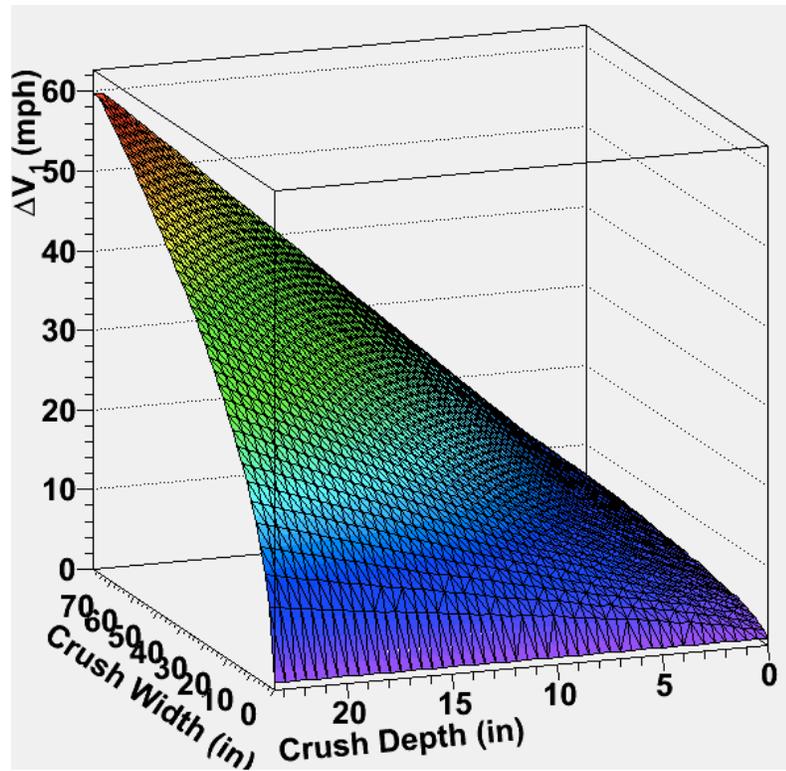

**Figure 10: Delta-V values displayed using surface plot feature of the "TGraph2D" class.**

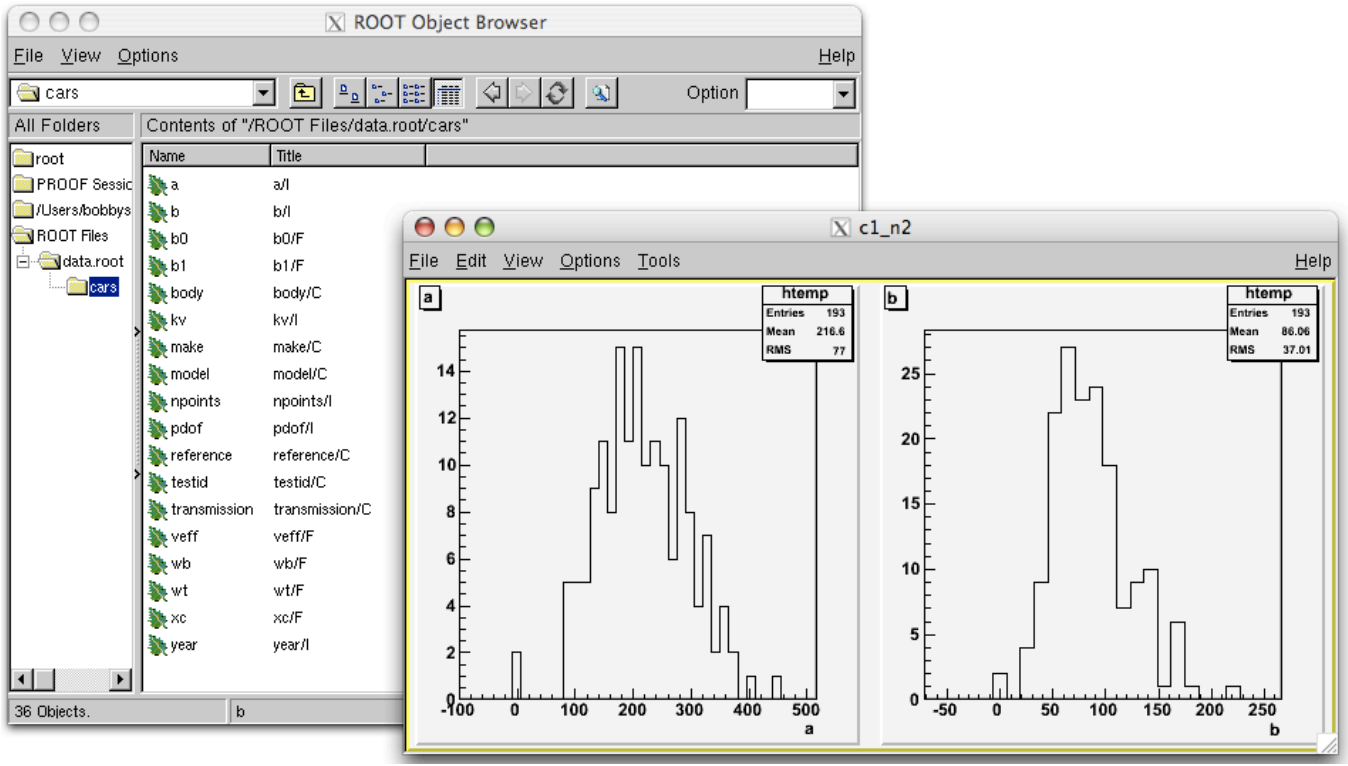

Figure 11: ROOT's object browser.